\documentstyle[preprint,eqsecnum,epsf,aps,floats,tighten]{revtex}
\input{psfig}
\def\D0{D\O}
\def\met{\mbox{${\hbox{$E$\kern-0.6em\lower-.1ex\hbox{/}}}_T$}} 
\def\d0draft{}
\def\etal{{\sl et al.}}
\begin{document}
%
%
\title{ \bf Search for $R$-parity Violating Supersymmetry in the
Dielectron Channel}

%
\author{                                                                      
B.~Abbott,$^{45}$                                                             
M.~Abolins,$^{42}$                                                            
V.~Abramov,$^{18}$                                                            
B.S.~Acharya,$^{11}$                                                          
I.~Adam,$^{44}$                                                               
D.L.~Adams,$^{54}$                                                            
M.~Adams,$^{28}$                                                              
S.~Ahn,$^{27}$                                                                
V.~Akimov,$^{16}$                                                             
G.A.~Alves,$^{2}$
% ...
\\                                                                            
\vskip 0.30cm                                                                 
\centerline{(D\O\ Collaboration)}                                             
\vskip 0.30cm 
}                                                                             
\address{                                                                     
\centerline{$^{1}$Universidad de Buenos Aires, Buenos Aires, Argentina}       
\centerline{$^{2}$LAFEX, Centro Brasileiro de Pesquisas F{\'\i}sicas,         
                  Rio de Janeiro, Brazil}                                     
\centerline{$^{3}$Universidade do Estado do Rio de Janeiro,                   
                  Rio de Janeiro, Brazil}                                     
\centerline{$^{4}$Institute of High Energy Physics, Beijing,                  
                  People's Republic of China}                                 
\centerline{$^{5}$Universidad de los Andes, Bogot\'{a}, Colombia}             
\centerline{$^{6}$Universidad San Francisco de Quito, Quito, Ecuador}         
\centerline{$^{7}$Institut des Sciences Nucl\'eaires, IN2P3-CNRS,             
                  Universite de Grenoble 1, Grenoble, France}                 
\centerline{$^{8}$DAPNIA/Service de Physique des Particules, CEA, Saclay,     
                  France}                                                     
\centerline{$^{9}$Panjab University, Chandigarh, India}                       
\centerline{$^{10}$Delhi University, Delhi, India}                            
\centerline{$^{11}$Tata Institute of Fundamental Research, Mumbai, India}     
\centerline{$^{12}$Kyungsung University, Pusan, Korea}                        
\centerline{$^{13}$Seoul National University, Seoul, Korea}                   
\centerline{$^{14}$CINVESTAV, Mexico City, Mexico}                            
\centerline{$^{15}$Institute of Nuclear Physics, Krak\'ow, Poland}            
\centerline{$^{16}$Institute for Theoretical and Experimental Physics,
                   Moscow, Russia}                                            
\centerline{$^{17}$Moscow State University, Moscow, Russia}                   
\centerline{$^{18}$Institute for High Energy Physics, Protvino, Russia}       
\centerline{$^{19}$Lancaster University, Lancaster, United Kingdom}           
\centerline{$^{20}$University of Arizona, Tucson, Arizona 85721}              
\centerline{$^{21}$Lawrence Berkeley National Laboratory and University of    
                   California, Berkeley, California 94720}                    
\centerline{$^{22}$University of California, Davis, California 95616}         
\centerline{$^{23}$University of California, Irvine, California 92697}        
\centerline{$^{24}$University of California, Riverside, California 92521}     
\centerline{$^{25}$Florida State University, Tallahassee, Florida 32306}      
\centerline{$^{26}$University of Hawaii, Honolulu, Hawaii 96822}              
\centerline{$^{27}$Fermi National Accelerator Laboratory, Batavia,            
                   Illinois 60510}                                            
\centerline{$^{28}$University of Illinois at Chicago, Chicago,                
                   Illinois 60607}                                            
\centerline{$^{29}$Northern Illinois University, DeKalb, Illinois 60115}      
\centerline{$^{30}$Northwestern University, Evanston, Illinois 60208}         
\centerline{$^{31}$Indiana University, Bloomington, Indiana 47405}            
\centerline{$^{32}$University of Notre Dame, Notre Dame, Indiana 46556}       
\centerline{$^{33}$Purdue University, West Lafayette, Indiana 47907}          
\centerline{$^{34}$Iowa State University, Ames, Iowa 50011}                   
\centerline{$^{35}$University of Kansas, Lawrence, Kansas 66045}
\centerline{$^{36}$Kansas State University, Manhattan, Kansas 66506}          
\centerline{$^{37}$Louisiana Tech University, Ruston, Louisiana 71272}        
\centerline{$^{38}$University of Maryland, College Park, Maryland 20742}      
\centerline{$^{39}$Boston University, Boston, Massachusetts 02215}            
\centerline{$^{40}$Northeastern University, Boston, Massachusetts 02115}      
\centerline{$^{41}$University of Michigan, Ann Arbor, Michigan 48109}         
\centerline{$^{42}$Michigan State University, East Lansing, Michigan 48824}   
\centerline{$^{43}$University of Nebraska, Lincoln, Nebraska 68588}           
\centerline{$^{44}$Columbia University, New York, New York 10027}             
\centerline{$^{45}$New York University, New York, New York 10003}             
\centerline{$^{46}$University of Rochester, Rochester, New York 14627}        
\centerline{$^{47}$State University of New York, Stony Brook,                 
                   New York 11794}                                            
\centerline{$^{48}$Brookhaven National Laboratory, Upton, New York 11973}     
\centerline{$^{49}$Langston University, Langston, Oklahoma 73050}             
\centerline{$^{50}$University of Oklahoma, Norman, Oklahoma 73019}            
\centerline{$^{51}$Brown University, Providence, Rhode Island 02912}          
\centerline{$^{52}$University of Texas, Arlington, Texas 76019}               
\centerline{$^{53}$Texas A\&M University, College Station, Texas 77843}       
\centerline{$^{54}$Rice University, Houston, Texas 77005}                     
}                                                                             
%


\maketitle
\begin{abstract}  
We report on a search for $R$-parity violating supersymmetry in
$p\bar{p}$ collisions at $\sqrt{s}$ = 1.8 TeV using the \D0 detector at 
Fermilab. Events with at least two electrons and four or more jets were 
studied. We observe 2 
events in $99\pm4.4$~pb$^{-1}$ of data, consistent with the expected 
background of 
$1.8\pm0.4$ events. This result is interpreted within the framework of minimal 
low-energy supergravity supersymmetry models. 
Squarks with mass below 243 GeV/$c^2$ and gluinos with mass below 
227 GeV/$c^2$ are excluded at the 95\% confidence level (C. L.) for $A_0=0$, 
$\mu\ <\ 0$, tan~$\beta=2$ and a finite value for any one of the six 
$R$-parity violating couplings ${\lambda{'}}_{1jk}$ ($j=$1, 2 and $k=$1, 2, 3).
\end{abstract}
\vspace{1cm}
\hspace{.6cm} The standard model (SM) has survived many precision tests. 
However, it is thought incomplete, and supersymmetry (SUSY)~\cite{susy} is 
considered an attractive 
extension to the SM because it protects the Higgs mass from large radiative 
corrections and can provide a dynamical means for breaking electroweak 
symmetry. SUSY predicts for each particle in the SM,
a partner with spin differing by half a unit. In its general form, the
theory contains over one hundred free parameters. For our comparison with 
data, we have therefore chosen the more tractable framework provided by
minimal low-energy supergravity (mSUGRA)~\cite{sugra}, which has 
only five free parameters: a common mass for scalars ($m_0$), 
a common mass for all gauginos ($m_{1/2}$), and a common trilinear coupling 
constant ($A_0$), all specified at the grand unification scale. The other two 
parameters are the ratio of the vacuum expectation values of the two Higgs 
doublets (tan~$\beta $), and the sign of the Higgsino mass parameter $\mu$.
The masses and couplings at the weak scale are obtained from these five 
parameters by solving a set of renormalization group equations.

Most of the searches for supersymmetric particles reported thus far have 
assumed the conservation of a multiplicative quantum number called 
$R$-parity~\cite{rparity}. 
$R$-parity is defined as $R=(-1)^{3B+L+2S}$, where $B$, $L$ and $S$ are the 
baryon, lepton and spin quantum numbers, respectively. $R$ is $+1$ for SM 
particles, and $-1$ for their SUSY partners. In SUSY, $R$-parity violation 
can occur quite naturally through the following Yukawa coupling terms in the 
superpotential:

\begin{center}
$\lambda_{ijk}L_{i}L_j{\overline{E}}_k +
 \lambda_{ijk}^{'}L_{i}Q_{j}{\overline{D}}_k +
 \lambda_{ijk}^{''}{\overline{U}}_i{\overline{D}}_j
 {\overline{D}}_k$
\end{center}

\noindent where $L$ and $Q$ are the SU(2)-doublet lepton and quark superfields;
$E$, $U$, and $D$ are the singlet lepton, up and down type quark superfields,
respectively; and $i$, $j$ and $k$ are the 
generation indices. The Yukawa couplings are antisymmetric in the same 
superfield indices. Thus, there can be up to 45 new Yukawa terms. 
We have therefore made the following simplifying assumptions 
for our analysis:

\begin{itemize}
\item Among the 45 $R$-parity violating couplings, only one dominates. 
This is motivated by the fact that the new couplings are similar 
to the SM Yukawa couplings, for which the top quark Yukawa term dominates.
Moreover, bounds on products of two couplings are generally  
stringent, because the presence of more than one coupling can induce 
rare processes such as flavor changing neutral currents at the tree 
level~\cite{product}.
\item The strength of the $R$-parity violating coupling under consideration 
is $> 10^{-3}$, so that the lightest supersymmetric particle (LSP) decays 
close to the interaction vertex. This is consistent with the existing upper 
bounds on the strength of the couplings from low energy 
experiments~\cite{bounds}.
\item The strength of the finite $R$-parity violating coupling term is 
significantly smaller than the gauge couplings. Thus, supersymmetric particles 
are produced in pairs, and $R$-parity violation manifests itself only in the 
decay of the LSP.  
\end{itemize}

Of the three kinds of Yukawa coupling terms, the $B$-violating ${\lambda}''$
are difficult to study at the Fermilab Tevatron as they lead to events with 
multiple jets that would  be overwhelmed by large backgrounds from QCD 
production of jets. However, the  $L$-violating $\lambda$ and ${\lambda}'$ 
couplings give rise to multilepton and multijet final 
states~\cite{cdf1}, which provide excellent signatures. 

\begin{figure}[h]
\centerline {\psfig{figure=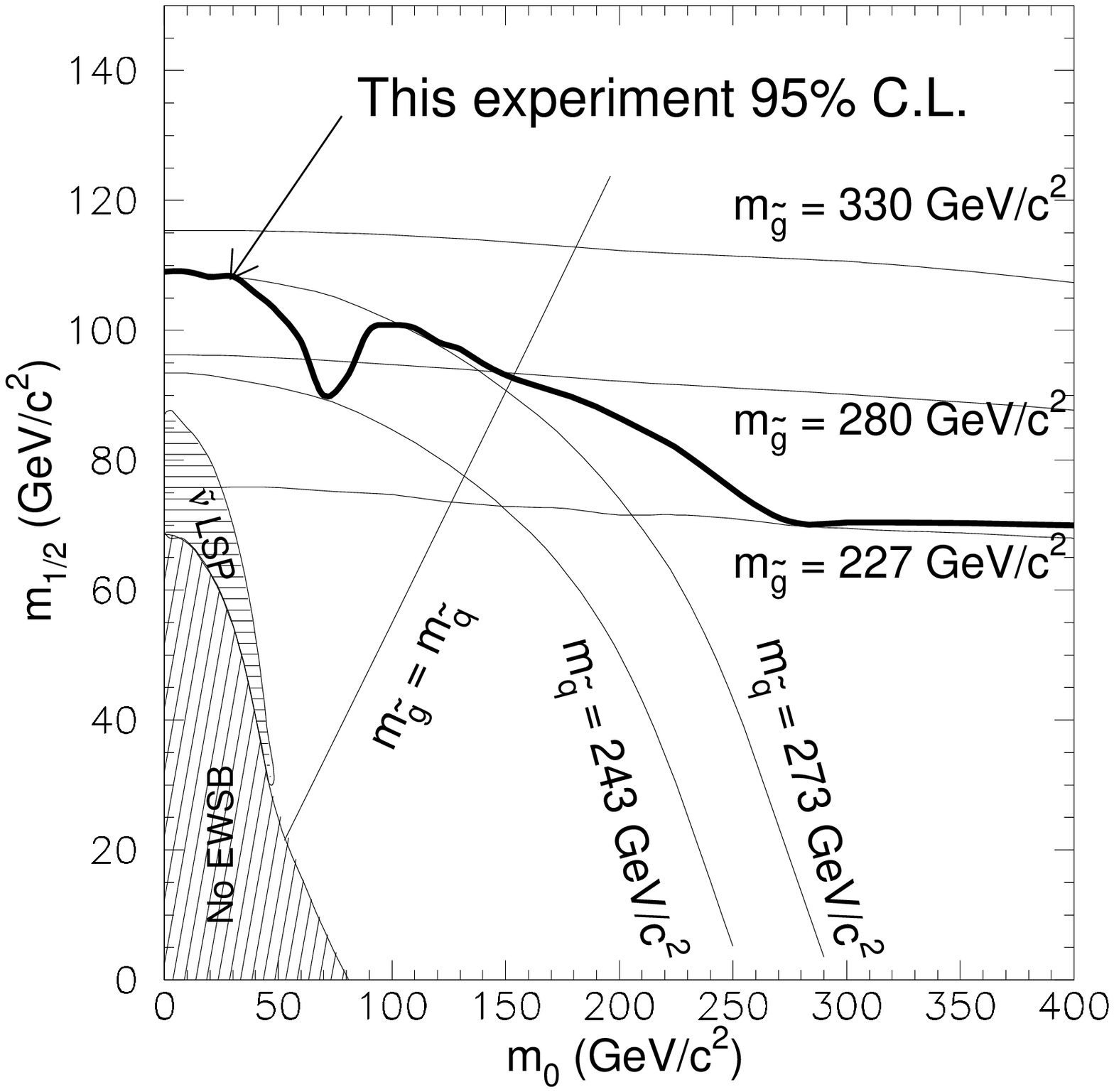,width=5.5in}}
\caption{
Exclusion contour in the $(m_0,m_{1/2})$ plane for $A_0=0$, $\mu\ < 0$, 
tan~$\beta \ =2$ and a finite ${\lambda}'_{1jk}$ ($j=$1, 2 
and $k=$1, 2, 3) coupling.
The region below the bold line is excluded at the 95\% C.L. The slanted 
hatched region is excluded for theoretical reasons. In the horizontally 
hatched region, the sneutrino is the LSP, but is excluded by searches at LEP  
(see the text).}
\label{fig:excl}
\end{figure}

This Letter reports on an analysis of the dielectron and four jets 
channel, interpreted in the mSUGRA framework, with $R$-parity violating decays 
of the LSP. In mSUGRA, the 
lightest neutralino is almost always the LSP except in a small region of the
($m_0, m_{1/2}$) plane where the sneutrino is the LSP (indicated in 
Fig.~\ref{fig:excl}). But the mass of the sneutrino in that region is below 
39 GeV/$c^2$ and, hence,  excluded 
($> 43.1$ GeV/$c^2$ at 95\% C.~L.)~\cite{sneutrino} by the known
invisible decay width of the $Z$ boson, assuming that there are three 
degenerate left handed sneutrino species.  We assume that all the 
$R$-parity violating couplings are small except for one of the six 
${\lambda}'_{1jk}$ ($j=$1, 2 and $k=$1, 2, 3), so that each LSP decays 
into one electron and two quarks which gives rise to final states 
with two or more electrons and four or more jets that we consider in our 
analysis.

The \D0 detector~\cite{det} has three major subsystems: a central tracker, an  
uranium liquid argon sampling calorimeter, and a muon spectrometer. 
Electrons 
are identified as narrow energy clusters that deposit more than 90\% of their
energy in the electromagnetic sections of the calorimeter. Jets are 
reconstructed using a cone algorithm~\cite{cone} with radius 0.5 in 
pseudorapidity $-$ azimuthal angle ($\eta,\phi$) space. The data used for this 
analysis were collected during the 1994--1995 Fermilab Tevatron
run at a center-of-mass energy of 1.8 TeV, and correspond to an
integrated luminosity of $99\pm4.4$~pb$^{-1}$~\cite{lum}.

Our initial sample of 163,140 events was collected with triggers requiring  
at least five calorimeter energy clusters, 
and $H_T \geq 115$ GeV, where $H_T$ is the scalar sum of the
transverse energies ($E_T$) of all calorimeter clusters.
In the offline analysis, we required at least two electrons, one with 
$E_T \geq 15$ GeV and the second with $E_T \geq 10$ GeV, and at least four 
jets with $E_T \geq 15$ GeV. Electrons had to be either within 
$|\eta| \leq 1.1$ (central calorimeter) or $1.5 \leq |\eta| \leq 2.5$ 
(forward calorimeters), to be isolated from other energy deposits, and to have
shower shape and tracking information consistent with that expected for  
electrons~\cite{thesis,run2} . Jets had to be within $|\eta| \leq 2.5$. 
The requirement on electrons reduced the original sample to just 38 events, 
and the subsequent requirements on jets reduced it further to 6 events.
To suppress backgrounds from electron decays of $Z$ bosons, 
we rejected events whose dielectron invariant mass was in the range of 
76--106 GeV/$c^2$. To ensure high trigger efficiency, events were further 
required to have $H_T > 150$ GeV. The cut on Z mass reduced our data sample 
to 2 events, but the $H_T$ requirement had no further impact.

The major inherent SM backgrounds are from Drell-Yan production (DY), from 
the decay of $t\overline{t}$ to electrons, and from the decay of $Z$ bosons to 
$\tau$ pairs that subsequently decay to electrons. Events arising from the 
misidentification of jets as electrons comprise the major source of 
instrumental background for this analysis. The huge reduction in our data 
sample from the requirement of having two isolated electrons reflects the 
fact that most of the events passing the trigger are due to QCD multijet 
production, and have no true isolated electrons.  

A {\sc geant}~\cite{geant} based simulation of the \D0 detector was used to 
estimate efficiencies of the kinematic cuts for non-instrumental backgrounds. 
Measured electron identification efficiencies were then folded in to calculate 
the net detection efficiency. Using $Z  (\rightarrow  ee)\ + $ jets data, we 
estimated single-electron identification efficiencies to be $0.68 \pm 0.07$ in
the central calorimeter, and $0.60 \pm 0.07$ in the forward calorimeters. 
{\sc isajet}~\cite{isajet} was used to generate DY events, with cross section 
increased by a factor of 1.7 to obtain agreement with the $Z$ + multijet data 
in the mass region of the $Z$ boson, yielding an expected $0.37 \pm 0.14$ 
(stat.) $\pm 0.14$ (syst.) events. Top quark events were generated using 
the {\sc herwig}~\cite{herwig} program. The measured cross section for 
$t\overline{t}$ production ($5.9\pm 1.7$ pb)~\cite{top} was used to estimate 
this contribution to background to be $0.07 \pm 0.02 \pm 0.02$ events. The 
production cross section of the $Z$ boson multiplied by its leptonic branching
fraction of ($221 \pm 11$) pb~\cite{lum} was used to estimate the background 
due to $Z  (\rightarrow  \tau \tau  \rightarrow  ee$) to be 
$0.07 \pm 0.01 \pm 0.02$ events. The instrumental background was estimated 
from data in two steps. First, from multijet data, we estimated the 
probability for misidentifying a jet as an isolated electron. This was
$(4.6 \pm 0.4) \times 10^{-4}$ in the central and 
$(1.4 \pm 0.2) \times 10^{-3}$ in the forward calorimeters. Within statistical
accuracy, these probabilities were found to be independent of $E_T$. We then 
selected a multijet data sample passing the same kinematic requirements as our
data sample, but requiring two additional jets instead of two electrons. The 
number of background events was estimated to be $1.27 \pm 0.24$  (with
negligible statistical uncertainty) by applying the probability for jet 
misidentification to these multijet data. The statistical components of 
uncertainty include fluctuations due to the finite sample size of simulated 
events and uncertainties in electron identification efficiencies. The 
systematic components of the uncertainty include those due to jet energy scale
and values of cross sections. Our two observed events are consistent with the 
expected background, both in the number of expected events 
$1.8 \pm 0.2 \pm 0.3$, and in their kinematic characteristics. In what 
follows, we interpret this null result in terms of an excluded region in  
mSUGRA parameter space. 

Using {\sc isajet}, we generated signal events at 125 points in the 
($m_0,m_{1/2}$) plane, with $A_0\ = 0$, $\mu \ < \ 0$ and tan~$\beta=2$. 
$R$-parity violating decays of the LSP are not available in {\sc isajet}. The 
desired decay modes and branching fractions for the LSP were therefore added 
separately. The branching fraction of the LSP into a charged lepton or 
neutrino depends on the gauge composition of the LSP, which in turn depends 
on the mSUGRA parameters. This was incorporated into {\sc isajet} using the 
calculation of Ref.~\cite{formula}. Once we specify a decay mode, {\sc isajet}
does a 3 body phase-space decay, but does not implement the appropriate 
matrix element into the differential distribution. The efficiency multiplied 
by the branching fraction for each signal sample was determined using a 
method similar to that used for the estimation of the SM background. The 
expected event yields in the ($m_0,m_{1/2}$) parameter space, corresponding 
to our integrated luminosity of 99~pb$^{-1}$, are given in 
Table~\ref{tab:signal}.

\begin{table}
\caption{Efficiency ($\epsilon$) multiplied by the branching fraction (B) and
the expected event yield $\langle N \rangle $, for several points in the 
($m_0,m_{1/2}$) 
parameter space. The uncertainties are the sum in quadrature of the 
statistical and systematic uncertainties (the statistical uncertainty 
dominates).}
\begin{tabular}{c|c|c|c} 
$m_0$ ( GeV/$c^2$ )  & $m_{1/2}$ ( GeV/$c^2$ ) &  $\epsilon B (\%)$  & 
$\langle N \rangle $ \\\hline
   0 & 120&$1.59 \pm0.23$&$  3.5\pm 0.5$ \\
  50 & 110&$1.49 \pm0.22$&$  2.8\pm 0.4$ \\
 120 & 110&$1.86 \pm0.25$&$  3.3\pm 0.4$ \\
 190 & 100&$1.56 \pm0.22$&$  3.4\pm 0.4$ \\
 280 &  90&$0.95 \pm0.15$&$  2.9\pm 0.4$ \\
 320 &  90&$0.71 \pm0.13$&$  2.2\pm 0.4$ \\
\end{tabular}
\label{tab:signal}
\end{table}

For each point in the ($m_0,m_{1/2}$) plane, we obtained a 95\% C. L. upper 
limit on the cross section for signal. This was done using a Bayesian 
technique, with a flat prior for the signal cross section, and Gaussian priors 
for the luminosity, efficiency, and expected background. The excluded region 
in the ($m_0,m_{1/2}$) plane was then obtained by comparing the limits on the 
measured cross section with the leading-order SUSY prediction given by 
{\sc isajet}. This is shown in Fig.~\ref{fig:excl}. 
The slanted hatched area in Fig.~1 indicates the region in which the model 
does not produce radiative electroweak symmetry breaking. In the low $m_0$ 
region ($m_0<150$~GeV/$c^2$), the dominant SUSY process that contributes to 
the signal is pair production of squarks. Hence, in this region, the exclusion
contour follows a squark mass contour ($m_{\tilde{q}} =273$ GeV/$c^2$). The 
dip in the contour for $m_0=60-80$ GeV/$c^2$ can be attributed to the fact 
that the two electrons can originate either from the decay of LSPs or from 
other SUSY particles. In about 60\% of the cases, both LSPs decay into 
electrons. However, electrons arising from the decay of LSPs may not always 
pass the $E_T$ cut. In such cases additional electrons arising from the decay 
of the second lightest neutralino (${\tilde{\chi}}^0_2$) can make the event 
pass the $E_T$
criterion. But for $m_0=60-80$~GeV/$c^2$, sneutrinos become
lighter than the ${\tilde{\chi}}^0_2$, and the decay of ${\tilde{\chi}}^0_2$ to
${\tilde{\chi}}^0_1$ and neutrinos
(${\tilde{\chi}}^0_2 \rightarrow \nu \tilde{\nu};\  \tilde{\nu} \rightarrow
{\tilde{\chi}}^0_1 \nu$) becomes dominant. This reduces the overall
branching fraction to dielectrons, resulting in the observed dip.

As $m_0$ increases, the sneutrino becomes heavier than ${\tilde{\chi}}^0_2$, 
and consequently the branching fraction of ${\tilde{\chi}}^0_2$ to neutrinos 
decreases, leading to an increase in the rate for the competing selectron 
channel, thereby enhancing the branching into the dielectron mode. (That is, 
when the ${\tilde{\chi}}^0_2$ decay proceeds through a virtual sneutrino, the 
decay through a virtual selectron becomes competitive.) 
The exclusion contour therefore moves up and again follows the 
273 GeV/$c^2$ squark mass curve until the intermediate $m_0$ region 
(150~GeV/$c^2 < m_0 < 280$~GeV/$c^2$), where processes such as the production 
of gluinos, ${\tilde{\chi}}_1^{\pm}$, and ${\tilde{\chi}}_2^0$, start becoming
important. The masses of these particles, as well as their production cross 
sections, do not change much with the increase of $m_0$. As a result, the 
exclusion contour in this region becomes less dependent on $m_0$. 

Finally, in the asymptotic region ($m_0 > 280$~GeV/$c^2$), 
production of squarks becomes insignificant, and the contour of exclusion 
becomes totally independent on $m_0$. In Fig.~\ref{fig:excl}, 
we have overlaid contours of fixed gluino mass and the average of the masses 
of the first two generations of squarks. 
Squarks with mass below 243 GeV/$c^2$ and gluinos below 227 GeV/$c^2$ are 
excluded for $A_0\ =\ 0$, $\mu \ <\ 0$, tan~$\beta\ =\ 2$, and a finite value 
($> 10^{-3}$) for any one of the six 
${\lambda}'_{1jk}$ ($j=$1, 2 and $k=$1, 2, 3) couplings. For 
equal 
mass squarks and gluinos, the corresponding limit is 277 GeV/$c^2$.

\begin{figure}[h]
\centerline{\psfig{figure=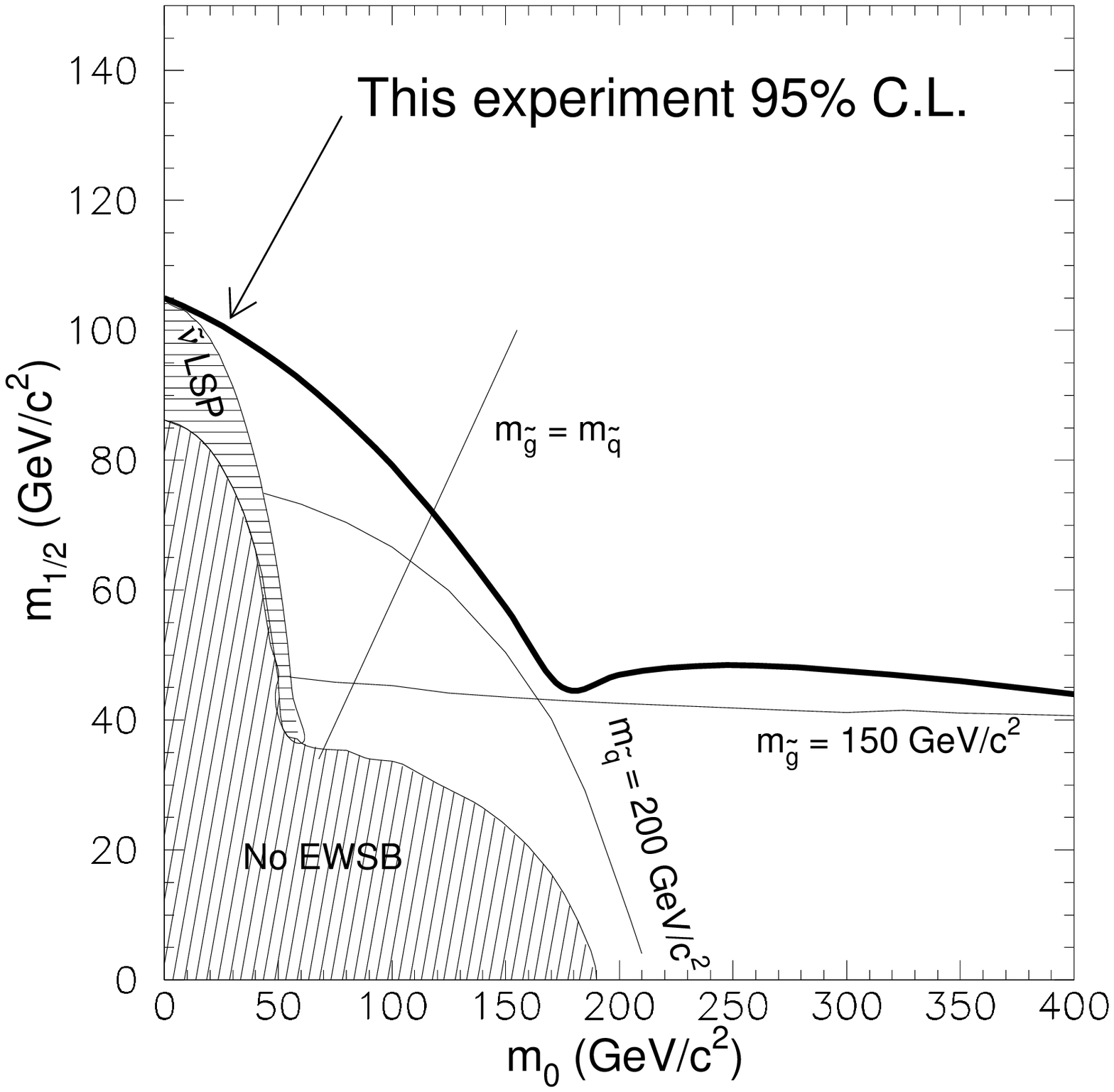,width=5.5in}}
\caption{
Exclusion contour in the $(m_0,m_{1/2})$ plane for 
$A_0=0$, $\mu\ < 0$, tan~$\beta \ = \ 6$, and a finite ${\lambda}'_{1jk}$ 
( $j=$1, 2 and $k=$1, 2, 3) coupling.}
\label{fig:tan6}
\end{figure}

We note that our results are essentially independent of the choice of $A_0$, 
as it affects only third generation sparticle masses. For $\mu > 0$ and higher
values of tan~$\beta$, the sensitivity of our search is expected to fall for 
two reasons: 1) the photino component of the LSP decreases, resulting in
the decrease of the branching fraction of the LSP into electrons; 2) the
charginos and neutralinos become light, resulting in events with softer 
electrons and jets that fail the kinematic requirements. 
We have estimated the sensitivity of our search for larger values of 
tan~$\beta$, by extrapolating our tan~$\beta\ =\ 2$ results using smeared
parton level {\sc isajet}~\cite{run2} (without full detector simulation). 
Figure 2 shows the region excluded at 95\% C. L. in the ($m_0,m_{1/2}$) plane 
for $A_0\ =\ 0$, $\mu \ <\ 0$, tan~$\beta\ =\ 6$, and a value $> 10^{-3}$ for 
any one of the six ${\lambda}'_{1jk}$ ($j=$1, 2 and $k=$1, 2, 3) couplings.
For higher values of tan~$\beta$, the sensitivity of this search 
deteriorates rapidly and requires a different analysis.

In conclusion, we have searched for events containing at least two electrons 
and four or more jets. Such events would be characteristic of processes 
involving the pair production of SUSY particles with the decay of 
the LSP through a $R$-parity violating coupling. Finding no excess of events 
beyond the prediction of the standard model, we interpret this result within 
the mSUGRA framework as an excluded region in the $(m_0,m_{1/2})$ plane for 
fixed values of $A_0$ and sign of $\mu $ and for several values of 
tan~$\beta$. This is the first result reported from Tevatron on search for 
$R$-parity violating SUSY involving several $\lambda'$ couplings in the 
mSUGRA framework. The Tevatron will continue to provide unique opportunities 
for searching for $R$-parity violating SUSY in a larger range of parameter 
space with the improved data anticipated from the next run~\cite{run2}. 

We thank the Fermilab and collaborating institution staffs for
contributions to this work and acknowledge support from the 
Department of Energy and National Science Foundation (USA),  
Commissariat  \` a L'Energie Atomique (France), 
Ministry for Science and Technology and Ministry for Atomic 
   Energy (Russia),
CAPES and CNPq (Brazil),
Departments of Atomic Energy and Science and Education (India),
Colciencias (Colombia),
CONACyT (Mexico),
Ministry of Education and KOSEF (Korea),
and CONICET and UBACyT (Argentina).

\end{document}